\documentclass[twocolumn]{aastex63}

\usepackage{amsmath}

\shorttitle{System Architecture and Planetary Obliquity}
\shortauthors{Vervoort et al.}

\graphicspath{{./}{figures/}}

\begin{document}
\title{System Architecture and Planetary Obliquity: Implications for Long-Term Habitability}

\author[0000-0001-9800-6723]{Pam Vervoort}
\affiliation{Department of Earth and Planetary Sciences, University of California, Riverside, CA 92521, USA}
\author[0000-0002-1160-7970]{Jonathan Horner}
\affiliation{Centre for Astrophysics, University of Southern Queensland, Toowoomba, QLD 4350, Australia}
\author[0000-0001-9800-6723]{Stephen R. Kane} 
\affiliation{Department of Earth and Planetary Sciences, University of California, Riverside, CA 92521, USA}
\author[0000-0002-3606-5071]{Sandra Kirtland Turner} 
\affiliation{Department of Earth and Planetary Sciences, University of California, Riverside, CA 92521, USA}
\author{James B. Gilmore} 
\affiliation{Australian Centre for Astrobiology, UNSW Australia, Sydney, New South Wales 2052, Australia}

\begin{abstract}

\noindent In the search for life beyond our Solar system, attention should be focused on those planets that have the potential to maintain habitable conditions over the prolonged periods of time needed for the emergence and expansion of life as we know it. The observable planetary architecture is one of the determinants for long-term habitability as it controls the orbital evolution and ultimately the stellar fluxes received by the planet. With an ensemble of \textit{n}-body simulations and obliquity models of hypothetical planetary systems, we demonstrate that the amplitude and period of eccentricity, obliquity, and precession cycles of an Earth-like planet are sensitive to the orbital characteristics of a giant companion planet. A series of transient, ocean-coupled climate simulations show how these characteristics of astronomical cycles are decisive for the evolving surface conditions and long-term fractional habitability relative to the modern Earth. The habitability of Earth-like planets increases with the eccentricity of a Jupiter-like companion, provided that the mean obliquity is sufficiently low to maintain temperate temperatures over large parts of its surface throughout the orbital year. A giant companion closer in results in shorter eccentricity cycles of an Earth-like planet but longer, high-amplitude, obliquity cycles. The period and amplitude of obliquity cycles can be estimated to first order from the orbital pathways calculated by the \textit{n}-body simulations. In the majority of simulations, obliquity amplitude relates directly to the orbital inclination whereas the period of obliquity cycles is a function of nodal precession and the proximity of a giant companion.

\end{abstract}



\section{Introduction} \label{sec:intro}

\noindent A wide range of factors determine whether or not an exoplanet may be considered habitable \citep[e.g.][]{HabRev,cockell2016}, although observational constraints often limit us to applying the 'Habitable Zone' (HZ) concept to evaluate planetary habitability. The HZ describes the region around a star in which an Earth-sized planet can theoretically host liquid surface water based on assumptions about the atmospheric properties and the incident stellar flux. One of the great advantages of the HZ concept is that the luminosity of a host star and the semi-major axis of an exoplanet, that both control the stellar flux, can typically be well constrained by observation \citep[e.g.][]{marcy2005}, allowing the HZ status of a given planet to be quickly and easily assessed.

However, it is crucial to take the temporal evolution of those variables into consideration when assessing a planet's long-term habitability potential \citep[e.g.][]{menou2003,spiegel2010,meadows2018,truitt2020}. In terms of planet habitability, attention should be focused on those planets that are likely to maintain temperate conditions over the prolonged periods of time relevant for the emergence and expansion of life as we know it. This requires a detailed investigation of the dynamic evolution of the planetary system. For instance, the luminosity of stars slowly increase during their time on the ``main sequence'' phase -- e.g. the Sun's luminosity increased by at least 30\% over the last four billion years \citep[][and references therein]{feulner}. The location of the HZ slowly moves outward on billion-year timescales as the stellar luminosity evolves \citep[]{rushby2013,truitt2020}.

The star-planet distance varies on much shorter timescales, particularly for planets on eccentric orbits. Eccentric planets experience great changes in the stellar flux as the planet moves along its orbit around the star \citep[]{spiegel2010,dressing2010,linsenmeier2015} and, in the most extreme cases, planets could leave and re-enter the traditional HZ through the course of an orbital year \citep[]{kane2012,williams2002,bolmont2016}. Whilst temporarily leaving the HZ does not necessarily make a planet inhospitable \citep[e.g.][]{dressing2010}, the stellar flux variability that results from such eccentric orbits will have consequences for the prevalent surface climate conditions that should be evaluated and quantified.

The distribution of the incoming stellar flux is a function of 1) the rotation rate, i.e. daytime vs nighttime duration, 2) the tilt (obliquity) of a planet's rotational axis relative to the star, and 3) the orientation (precession) of the axial tilt during a given time of year \citep{milankovitch1941,berger1978,dobrovolskis2013,linsenmeier2015}. Hemispheric differences in the incident stellar flux arise from precession. On eccentric planets, the hemisphere pointing to the star during perihelion receives greater stellar flux in summer than the other. Obliquity regulates the seasonal intensity as the axial tilt controls the stellar flux at a given latitude throughout the orbital year. The influence of obliquity can dominate over the influence of eccentricity depending on the degree of tilt and orbital ellipticity \citep{kane2017}. Ultimately, eccentricity in tandem with obliquity and precession control the incoming stellar flux and regulate the spatiotemporal surface conditions.

Several studies have demonstrated the impact of obliquity on surface climate \citep[e.g.][]{spiegel2009,ferreira2014,linsenmeier2015,kilic2017,colose2019} and found that multiple stable climate states (ice-free, partially ice-covered, fully ice-covered) can exists under a range of obliquity values, assuming Earth-like atmospheric composition and rotation rates \citep[]{kilic2017}. Yet, obliquity and precession are often ignored when assessing planetary habitability potential because these astronomical parameters are unattainable with the current observational technology \citep{gaidos2004}.

An additional complication comes from the fact that planets are rarely solitary. Gravitational interactions between planets drive cyclic variations in orbital eccentricity and inclination on timescales of thousands to millions of years that, in turn, perturb the axial obliquity and precession \citep[e.g.][]{kinoshita1977,laskar1993,atobe2004}. The extremity and duration of these quasi-periodic orbital and rotational cycles control the surface conditions and can significantly influence the potential habitability of a planet in the long-term \citep[]{spiegel2010,armstrong2014,way2017,deitrick2018b}. For Earth, variations in its orbital eccentricity, mainly induced by Jupiter's gravitational influence, are relatively small ($\sim$5$\%$) \citep{laskar1993}. Nonetheless, those modest variations in combination with obliquity and precession dynamics, have regulated the growth and retreat of polar ice caps across the Quaternary glacial-interglacial cycles. Large parts of the Earth's surface fortunately remained ice-free (i.e. habitable) even during the most extreme glacial cycles of the recent geologic past. Astronomical cycles on other HZ exoplanets may, however, be more extreme and cause those planets to experience variations in stellar flux of different magnitude and frequency that could repeatedly increase or decrease its habitability \citep{spiegel2010,armstrong2014}.

\begin{figure*}[ht!]
\epsscale{0.80}
\plotone{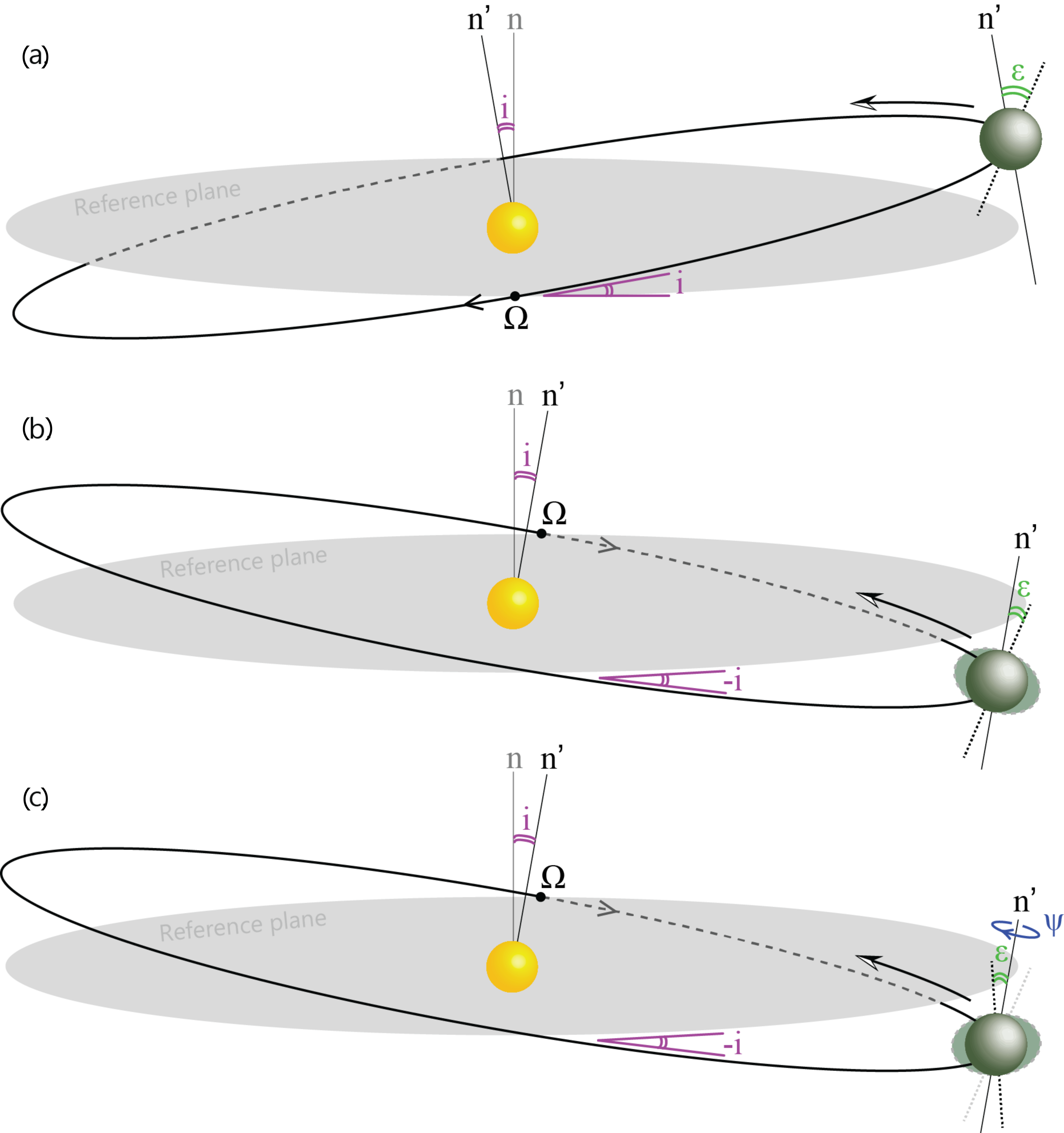}
\caption{Schematic depiction of the impact of nodal precession $\Omega$ on the obliquity $\epsilon$ of a planet. (a) The orbit is tilted $i\degr$ relative to the reference plane with orbital normal $n$. The Northern Hemisphere of the planet is tilted away during a given time of year with an angle $\epsilon$ between the rotational axis and the orbit normal $n'$. (b) The nodal precession has rotated the orbital plane 180\degr{} and the orbit is tilted $-i\degr$ relative to the reference plane. If the rotational axis would remain fixed relative to the fixed background stars, the new obliquity is $\epsilon-2i$. (c) However, solar and lunar torques pull the equatorial bulge toward the solar and lunar planes, driving precession of the rotational axis while affecting the obliquity.  \label{fig:nodal}}
\end{figure*}

Despite the observational challenges that limit full characterization of HZ exoplanets and the planetary architecture of their host systems, the use of $n$-body simulations in combination with obliquity and climate models can help to identify the HZ planets with the greatest long-term habitability potential. $N$-body simulations are now regularly used to evaluate the dynamical stability of planetary systems, refine orbital parameters, and reconstruct the orbital evolution of newly discovered exoplanets on timescales of millions of years \citep[e.g.][]{HR8799,marshall2020,HUAqr,horner2019orbital,matsumura2013,toth2014,agnew2019,kane2021}. An obvious extension of such modelling is to apply the orbital results to obliquity models to calculate a range of possible obliquity solutions based on assumptions about the planetary characteristics \citep[e.g.][]{shan2018,quarles2020,quarles2022}. Finally, the eccentricity, obliquity, and precession parameters can be applied to climate models to simulate the long-term climate cycles of the given planet. This sequential methodology has been applied to both hypothetical \citep{armstrong2014} and observed \citep{shields2016,quarles2020,quarles2022} planetary systems.

\begin{figure*}[ht!]
\plotone{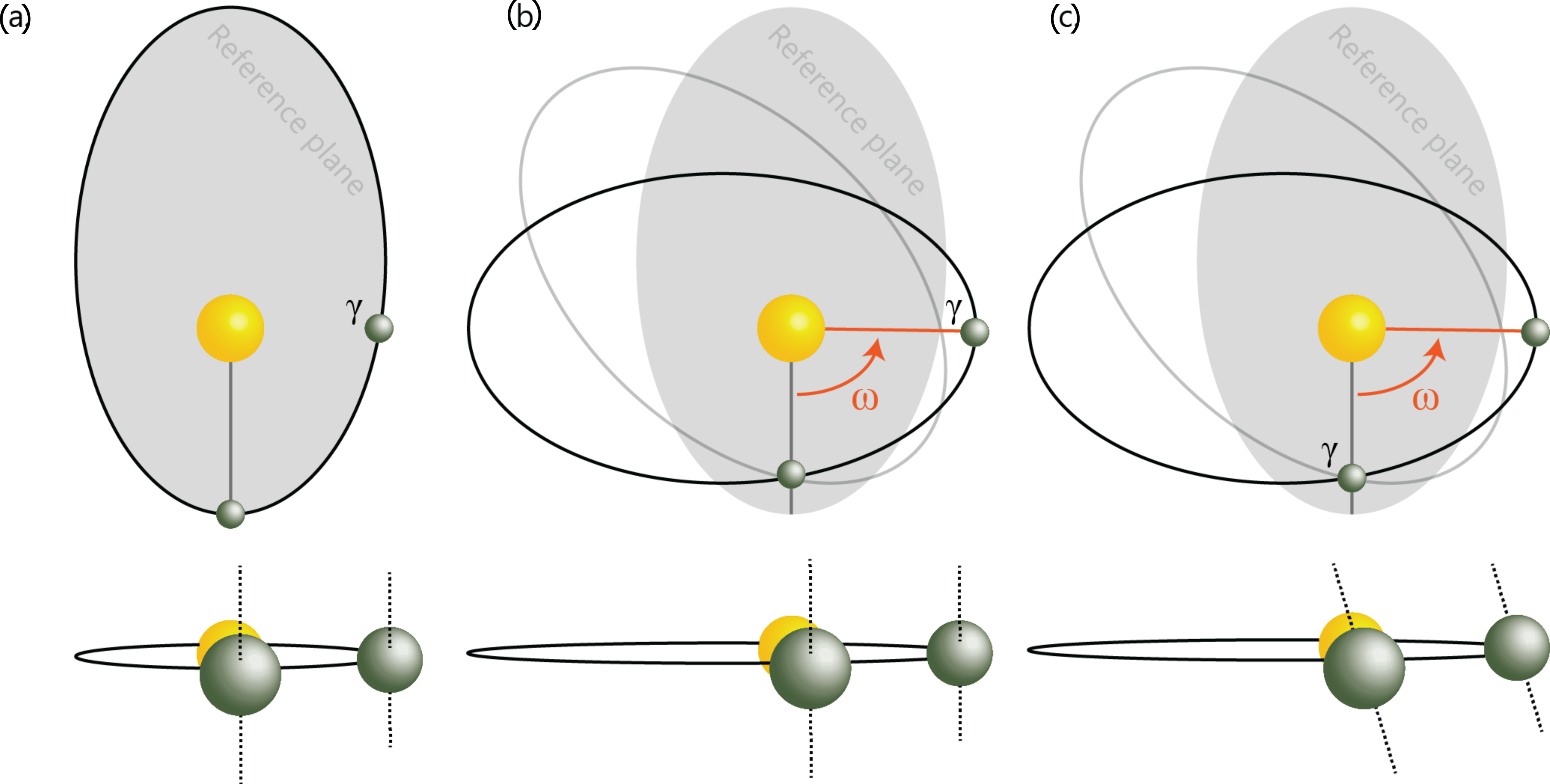}
\caption{Schematic depiction of the impact of apsidal and axial precession on the planet's climatic precession. Upper panels show top-views. Bottom panels show side-views. (a) A planet on a plane of reference with the Southern Hemisphere tilted to the Star during perihelion and the position of the vernal equinox indicated with $\Upsilon$. (b) The orbit has rotated relative to the fixed background stars, a movement known as apsidal precession $\omega$. If the rotational axis would remain fixed relative to the background stars, the vernal equinox now occurs during perihelion. (c) However, the planetary axis rotates over time so the location of the vernal equinox moves relative to the reference direction. The apsidal precession direction is opposite from the axial precession thus climatic precession cycles experienced by the planet are more rapid than the axial precession cycles.
\label{fig:apsidal}}
\end{figure*}

The simplest simulations of exoplanet climates feature steady state scenarios that generate surface climate conditions under a fixed astronomical configuration, providing first order estimates of how surface climate conditions might change with astronomical forcing \citep{williams2003,spiegel2009,dressing2010,dobrovolskis2013,ferreira2014,linsenmeier2015,wang2016,kilic2017,kang2019}. However, climate systems do not transition instantaneously from one stable state into another, but instead experience some form of climate inertia and require time to adjust to new conditions. Slowly responding ice sheets, vast ocean basins with a large heat capacity, or slow (bio)geochemical processes can make climate systems resistant to rapid changes in the radiative energy balance. Transient climate simulations that include processes contributing to climate inertia are therefore necessary to study the long-term astronomical climate evolution that results from the dynamical evolution of a planetary system \citep{spiegel2010,armstrong2014,way2017,deitrick2018a,georgakarakos2018}.

In this study, we use a $n$-body ensemble of hypothetical planetary systems \citep{horner2019} and apply obliquity calculations to investigate the extent to which planetary architecture influences the long-term dynamical evolution of an Earth-like planet in the HZ. Transient ocean-coupled climate models, with the three astronomical parameters varying simultaneously, are applied to a subset of these across the million-year timescale relevant for the evolutionary development of life, to evaluate their impact on the long-term habitability potential. 

The methods are outlined in Section \ref{sec:methods}. Our results are presented in Section \ref{sec:results}, following which, we discuss the implications for the long-term climate evolution and habitability in Section \ref{sec:climate}. Two parameters important for the reconstruction of obliquity cycles, but uncertain for most HZ planets in other exoplanet systems, are the dynamical ellipticity and tidal torques. Their impact on spin dynamics are addressed in Section \ref{sec:discussion} before summarizing our main findings in Section \ref{sec:conclusion}.

\section{Methods} \label{sec:methods}

\noindent To illustrate to what extent planetary architecture and orbital dynamics influence the long-term habitability potential of planets in the Habitable Zone, we use an ensemble of pre-existing hypothetical planetary systems very similar to our own Solar system \citep{horner2019}. The planetary architecture is modified to create hypothetical systems via systematic changes to the initial orbital characteristics of the Jupiter-mass planet. 

The resulting orbital evolution of the Earth-like planet is recorded and used as input to the obliquity model to calculate its obliquity and precession cycles. Finally, transient climate simulations are carried out to estimate the impact of the eccentricity, obliquity, and precession on the surface climate. The results are reported in terms of areal sea ice extent and fractional habitability (as per \cite{spiegel2009}), allowing us to assess the impact of planetary architecture on the long-term habitability potential. For direct comparability between the habitability potential of our planet in its current state and that of an Earth-like planet in a planetary system with a different architecture, we adopt Earth-like parameter values in the obliquity and climate models.

\subsection{Dynamical n-body simulations} \label{subsec:$n$-body}

\noindent A detailed description of the $n$-body framework is provided in \cite{horner2019}. In summary, an extensive suite of $n$-body simulations was carried out using the Hybrid integrator within the {\sc Mercury} software \citep{chambers1999}. The software was modified to take account of first-order post-Newtonian corrections \citep{gilmore08} to ensure that the orbital evolution of the planet Mercury, and its resulting perturbations on the other planets, is accurately described. The ensemble consists of 159,201 individual hypothetical (`alternative Solar system') simulations in which the initial orbits of Mercury, Venus, Earth, Mars, Saturn, Uranus, and Neptune were consistent with the modern Solar system configuration and Jupiter's initial orbit was altered. Our hypothetical systems are not tied to any prior planet formation modelling. Rather than presenting the final results of a planet formation process, they are instead used as theoretical test cases to illustrate how different planetary architectures would impact the orbital variability of an Earth-like planet - using the modern Earth as a convenient and well-studied example.

In the ensemble of 399$\times$399 simulations, the initial Jupiter semi-major axis~($a_{J}$) ranges from 3.2 to 7.2~au, within the 3-10~au range of peak-frequency proposed for gas giants \citep{bryan2016}. The initial eccentricity of Jupiter ($e_{J}$) ranges from 0.0 to 0.4, consistent with the observed range of eccentricities in multi-planet systems \citep[e.g.][]{weiss2013,Motalebi2015,eylen2015}. It is likely that Jupiter-like planets at orbital radii less than 3.5~au would limit the formation of terrestrial planets beyond 1~au, whilst a Jupiter as the inner most giant at greater than 10~au orbital radii would allow formation of additional planets beyond Mars' current semi-major axis and would thus likely result planetary architectures very different than our own Solar system \citep{nagasawa2007}. 

Stable simulations were integrated over 10~Myr. Approximately 74$\%$ of the simulations were deemed unstable and terminated prematurely when any of the planets collided with the Sun, with each other, or reached a heliocentric distance of 40~au.

\subsection{Obliquity model} \label{subsec:spinmodel}

\noindent We use the spin dynamics model described in \cite{laskar1993}, based on the equations of the rigid-Earth theory \citep{kinoshita1977}, to calculate the obliquity and climatic precession cycles of an Earth-like planet and their evolution over time (see Appendix and equations therein). The model is calibrated specifically to Earth for which the spin dynamics are, to the first order, controlled by orbital dynamics and tidal interaction with the Moon and the Sun \citep{ward1982,laskar1993,laskar1993moon,williams1993,neron1997,waltham2015}. Other planet dependent parameters, including Earth's angular momentum ($\nu$  = 2$\pi$ 24h$^{-1}$) and dynamical ellipticity ($E_D$ = 0.00328) are taken to be equal to the canonical values for the modern day Earth. An initial obliquity of $\epsilon_0$ = 23.4$\degr$ is chosen to maintain consistency across all experiments, but we note that the solution of any obliquity model will be sensitive to the chosen initial conditions as shown in detail for a moonless Earth by \cite{lissauer2012}. A sensitivity analysis is conducted to verify our results (see Appendix).

The obliquity model is integrated for 10 Myr of which the last million years are simulated with the climate model. The relatively short integration time provides a lower estimate for the total range of obliquity variations. The total range of variation on 100 Myr-timescales likely differs by a few degrees \citep{lissauer2012}. Planetary system chaos may induce a shift in the mean obliquity on Gyr timescales that we are unable to capture, however, when observing a planet and assessing its long-term habitability potential in search for life, the most recent past and future ($\pm$500 Myr) obliquity variations are most relevant.

\begin{figure*}[ht!]
\plotone{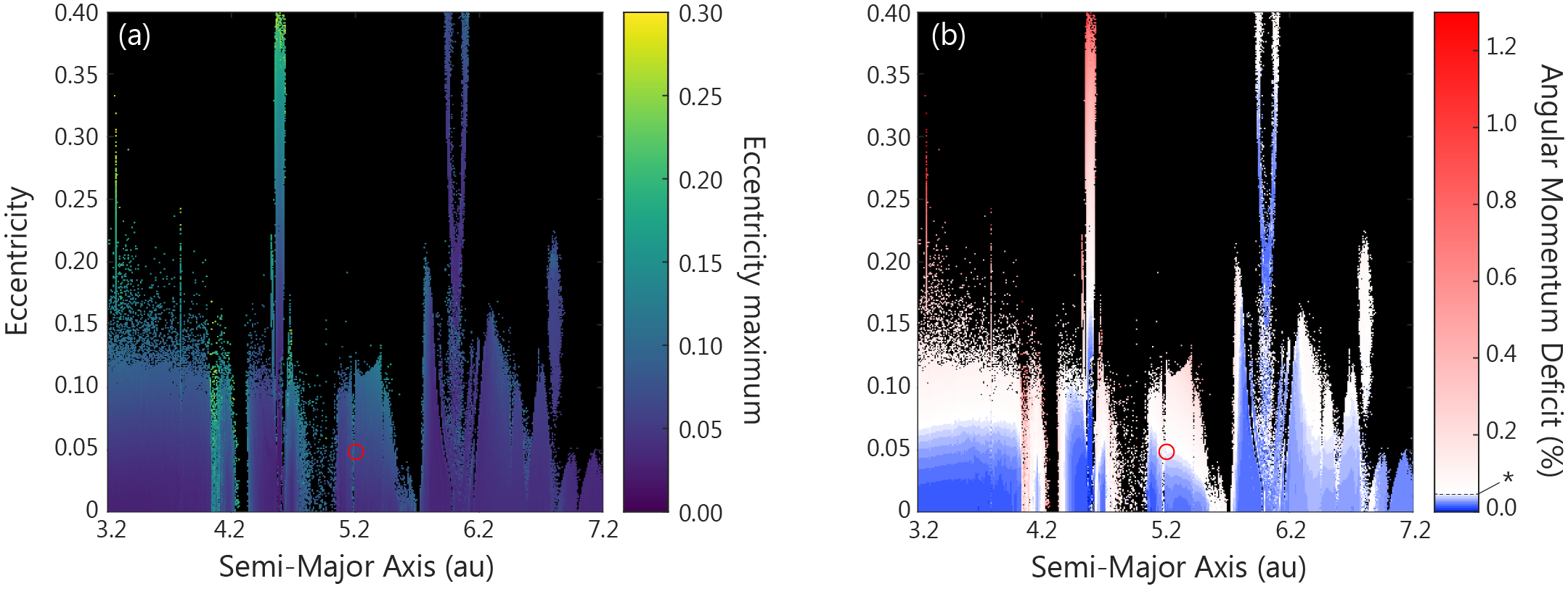}
\caption{Maximum eccentricity and mean angular momentum deficit. (a) The maximum eccentricity of an Earth-like planet under different planetary architectures, i.e. with a Jupiter-like planet having different initial semi-major axes and eccentricity. (b) The percentage of deviation in angular momentum (AMD) relative to that of a circular orbit, calculated as the mean deviation across the 10 Myr integration time. Blue colors indicate AMDs less than the modern Earth, below 0.0477\%. The red circle is the current position of Jupiter in our Solar system.
\label{fig:eccmax_AMD}}
\end{figure*}

A caveat of this obliquity model is that it does not account for any changes in the angular momentum of the Earth-Sun and Earth-Moon system. Angular momentum scales directly with eccentricity, and any major deviations in the orbital eccentricities of the Earth and Moon may therefore impact the results. However, the model has previously been used to reconstruct the Earth's obliquity and precession cycles for the past 10 million years and beyond within reasonable error \citep{laskar1993}, and is expected to produce accurate results in $>$82\% of simulations in which the angular momentum of the Earth-like planets is similar to that of the modern Earth (Figure \ref{fig:eccmax_AMD}b, blue-white shading). Hypothetical systems with more extreme eccentricity variations should be interpreted with caution as the deviation in angular momentum relative to its circular counterpart grows (Figure \ref{fig:eccmax_AMD}b, red shading) \citep[][]{barnes2004,kane2014}.

A near circular orbit of the Moon is assumed in all simulations to allow for direct comparability with our Earth. Whilst studies have suggested potential higher lunar eccentricities during the early evolution of the Earth-Moon system \citep[e.g.][]{touma1998}, tidal dissipation has circularized the lunar orbit over time \citep{zahnle2015}. Periodically elevated eccentricities of Earth may inject eccentricity into the lunar orbits that the model does not account for.

\subsection{Obliquity-precession dynamics} \label{subsec:obliquitybasics}

\noindent Gravitational interactions between Earth and the other planets in the Solar system cause quasi-periodic variability in Earth's eccentricity $e$, orbital inclination $i$, the longitude of the ascending node $\Omega$ (the angle between a reference direction and the ascending node, measured in a reference plane), and the argument of perihelion $\omega$ (the angle between the ascending node and perihelion, measured in the orbital plane). Such orbital cycles directly translate into systematic perturbations of the climatic precession $e \sin \varpi$ and obliquity $\epsilon$ experienced by the Earth. To emphasize: inclination is defined throughout the text as the angle between the orbital plane and a reference plane whereas obliquity refers to the angle between the planetary orbit normal and the rotational axis (Figure \ref{fig:nodal}).

Nodal precession describes the change in the orientation of the orbit normal relative to a reference plane. In other words, the tilt of the orbit relative to the fixed background stars changes over the course of a nodal precession cycle. A positive change reflects a clockwise motion when looking down at the system from above its north pole, as currently experienced by Earth. Planets in other systems can rotate anti-clockwise (negative change). To illustrate the effect of nodal precession on the planetary obliquity, assume for simplicity that the axial orientation remains fixed relative to the background stars. At t$_0$, Earth's obliquity is $\epsilon_0$ (Figure \ref{fig:nodal}a). After half a nodal precession cycle the orbital inclination has rotated by 180\degr, and the axial obliquity would reduce by twice the orbital inclination angle ($\epsilon_0-2i$) (Figure \ref{fig:nodal}b). 

However, the Earth is markedly oblate as a result of its relatively fast rotation. The orientation of the planet's spin axis rotates when solar and lunar torques essentially pull the equatorial bulge toward the ecliptic. These forces have been proposed to stabilize the long term obliquity oscillations of our planet \citep[e.g.][]{ward1982,laskar1993moon,williams1993}, although more recent work suggests that obliquity variations would still remain relatively sedate in absence of a large moon \citep{lissauer2012,li2014}. Because the planet's rotational axis as well as the orbit normal both precess in a clockwise direction (top-down view), the location of the equinoxes rotate. The equinoxes are defined by the two locations in a planetary orbit where the Sun is positioned directly above the equator (Figure \ref{fig:nodal}c). The nodal and axial precession of a planet combine to drive the axial obliquity cycles relative to the orbital plane. The full equations are shown in the Appendix (Equation A.2). To first order, the period of obliquity cycles can be approximated by:

\begin{equation}
\frac{1}{P_{\psi}} - \frac{1}{P_{\Omega}} = \frac{1}{P_{\epsilon}}
\end{equation}

\noindent where P$_{\psi}$ and ${P_{\Omega}}$ are the periods of axial and nodal precession, respectively. ${P_{\epsilon}}$ is the resulting period of the axial obliquity cycle. For the modern Earth (in yr), 

\[ \frac{1}{25,700} - \frac{1}{68,000} \approx \frac{1}{41,000} \] 

Apsidal precession describes the rotation of the orbital plane of a planet relative to the fixed background stars and, together with axial precession, determines the period of the climatic precession cycles (Figure \ref{fig:apsidal}). The climatic precession $\varpi$ is the change in the angle between the moving vernal equinox and perihelion, modulated by the orbital eccentricity following $e \sin \varpi$. When the orientation of the orbital plane is fixed, the vernal equinox rotates (clockwise) with an angular velocity equal to the axial precession (Figure \ref{fig:apsidal}b). However, the orbital plane also rotates over time, in the opposite direction for modern Earth, and thus the vernal equinox occurs slightly earlier (Figure \ref{fig:apsidal}c). The full equations are shown in the Appendix (Equation A.1). To first order, the period of climatic precession cycles can be approximated by:

\begin{equation}
\frac{1}{P_{\psi}} - \frac{1}{P_{\omega}} = \frac{1}{P_{\varpi}}
\end{equation}

\noindent where ${P_{\omega}}$ and is the period of apsidal precession, and ${P_{\varpi}}$ is the resulting period of the climatic precession cycle. For the modern Earth (in yr), 

\[ \frac{1}{25,700} - \frac{1}{-303,000} \approx \frac{1}{23,000} \]

\subsection{Atmosphere-ocean-sea ice model} \label{subsec:genie}
\noindent The simultaneous impact of the alternate eccentricity, obliquity, and climatic precession cycles on the surface temperature and sea ice extent are simulated with the GENIE Earth system model comprised of a coupled 2D Energy Moisture Balance Model (EMBM), 3D frictional geostrophic ocean model \citep{edwards2005,marsh2011}, and a thermodynamic sea ice model \citep{weaver2001}. The model, coupled to models of atmospheric chemistry and marine biogeochemistry, has previously been used to successfully simulate recent glacial and interglacial conditions \citep{marsh2006,ma2014,kemppinen2019}, to reconstruct a wide variety of other paleoclimate states throughout Earth's geologic history \citep{donnadieu2006,meyer2008,panchuk2008,crichton2020}, for comparison between steady state and transient runs \citep{lunt2006}, and to simulate changing surface conditions over a multi-million-year timescale in response to astronomical forcing \citep{vervoort2021}.  

Heat exchange between the ocean, sea ice, atmosphere, and the continents is balanced by incoming short wave radiation, sensible and latent heat fluxes, and outgoing long wave radiation calculated from the atmospheric relative humidity and greenhouse gas concentrations \citep{edwards2005}. We use a seasonally resolved set-up with 96 time steps per year and apply the modern solar constant of 1368~Wm$^{-2}$. The distribution of the incoming radiation is calculated directly from the astronomical parameters \citep[e.g.][]{berger1978}, which are 
specified in 1000~year intervals. The incoming short wave radiation is also a function of the prescribed surface albedo and temperature-dependent sea ice albedo. Atmospheric heat diffusivity is given by an exponential function that includes latitude as an input variable \citep{edwards2005}. Sea ice growth rates are calculated from the ice thickness, areal fraction, heat fluxes from the overlying atmosphere and underlying ocean, the density for water and ice, and the latent heat of fusion of ice \citep{weaver2001}.

For the purpose of this study, and to facilitate direct comparison with the present-day Earth, we adopt a planetary rotation rate of 2$\pi$ 24h$^{-1}$ and use a continental distribution similar to the Earth's modern configuration on a 36 by 36 equal-area grid (10$\degr$ longitude and uniform in sine of latitude). Geostrophic ocean dynamics are simulated spatially and across eight vertical ocean depth levels with a maximum depth of 5000~meters. Atmospheric CO$_2$ concentrations are prescribed to 278~ppm to recreate preindustrial greenhouse gas forcing. When implementing modern (fixed) astronomical input parameters, the simulated global mean air temperature is about 13$\degr$C, and the sea ice area is approximately 10$\times10^{6}$ km$^2$, consistent with the preindustrial surface of the Earth \citep{peng2018}. 

\begin{figure}[t!]
\plotone{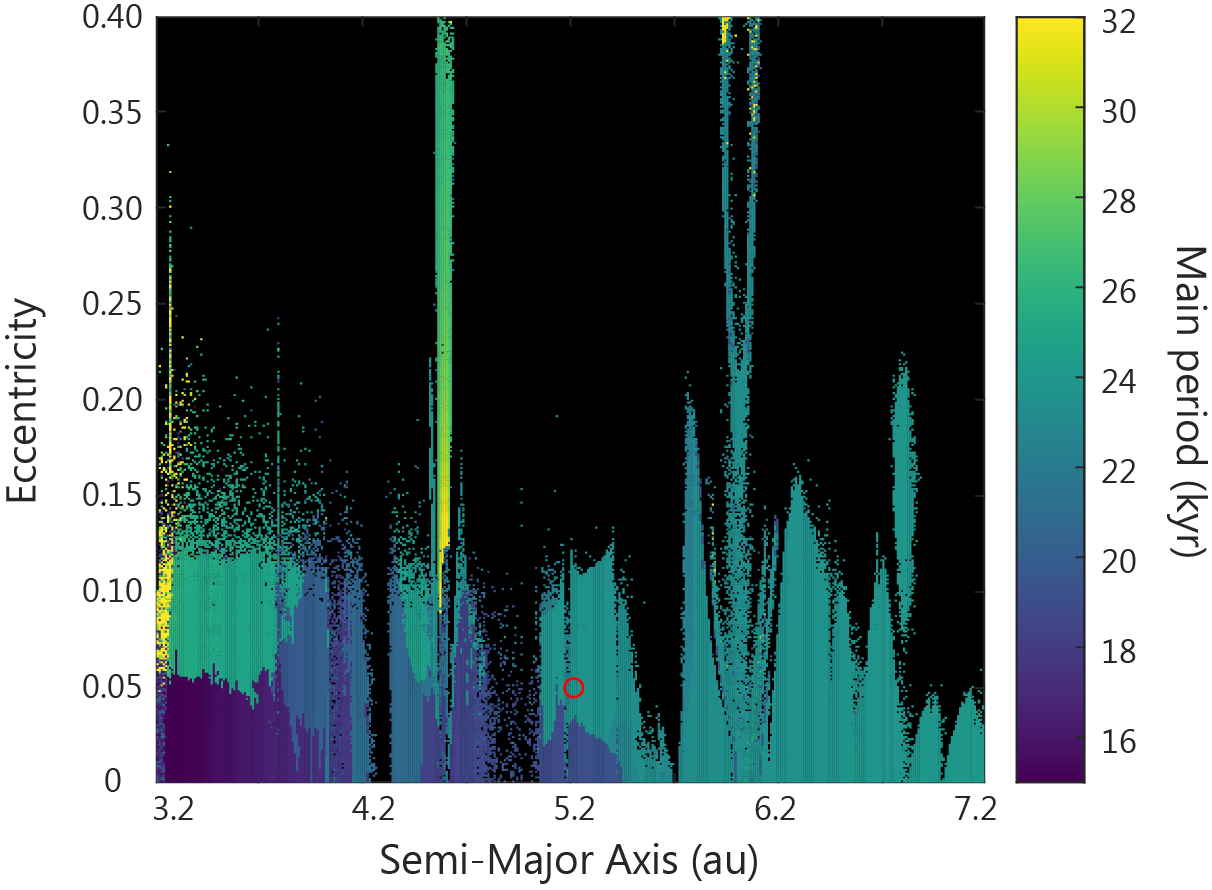}
\caption{The period of the main climatic precession cycle of an Earth-like planet under different planetary architectures, i.e. with a Jupiter-like planet having different initial semi-major axes and eccentricity. Climatic precession is defined here as $e \sin \varpi$. The red circle indicates the current position of Jupiter in our Solar system.
\label{fig:precessionperiods}}
\end{figure}

\begin{figure}[t!]
\plotone{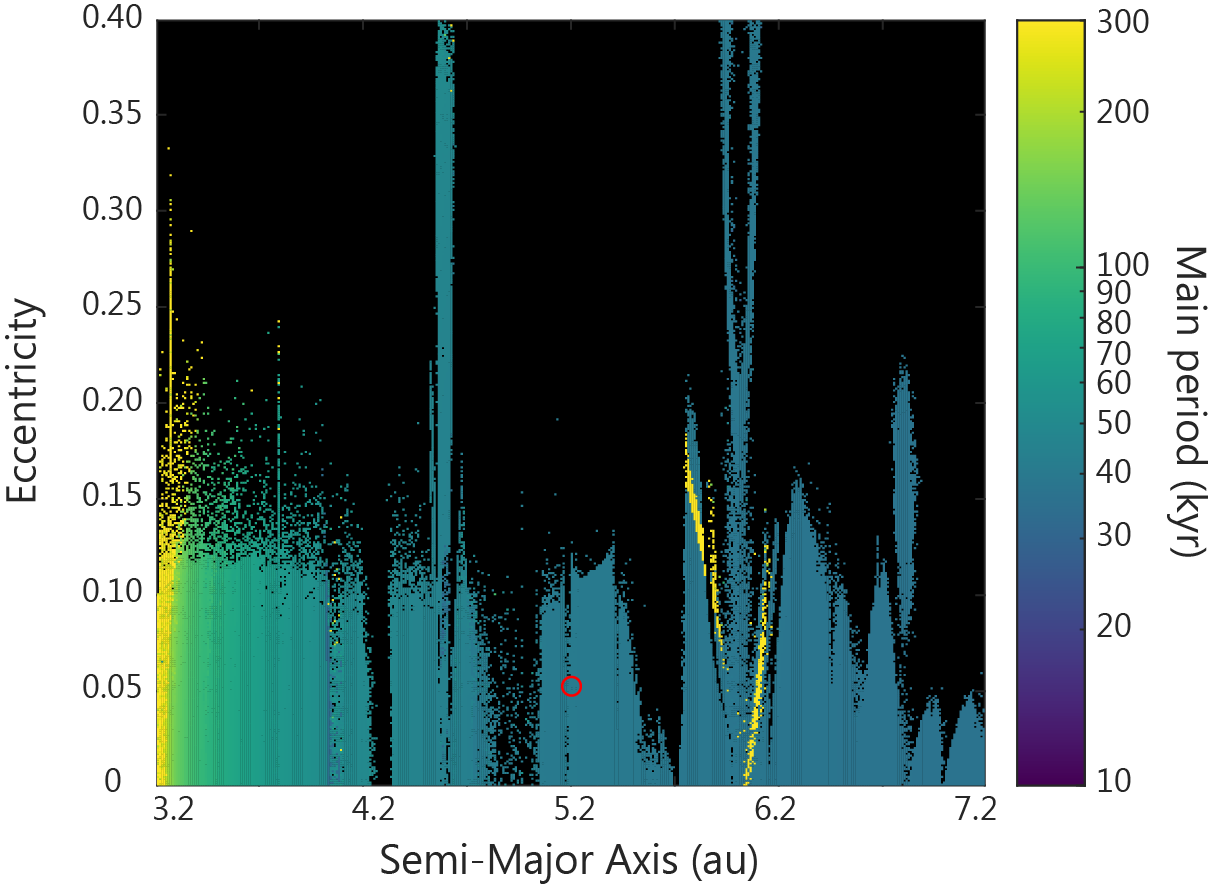}
\caption{The period of the main obliquity cycle of an Earth-like planet plotted on a logarithmic scale, under different planetary architectures, i.e. with a Jupiter-like planet having different initial semi-major axes and eccentricity. The red circle indicates the current position of Jupiter in our Solar system.
\label{fig:obliquityperiods}}
\end{figure}

\section{Results} \label{sec:results}

\noindent The dynamical results of the hypothetical $n$-body systems are published in \cite{horner2019}. To summarize, the ensemble shows that an eccentric Jupiter-mass planet in close proximity to an inner Earth-like planet results in more rapid eccentricity and inclination cycles of the Earth-like planet. Likewise, a Jupiter-mass planet that is located at greater heliocentric distance, or on a more circular orbit, drives low frequency orbital cycles of the Earth-like planet. The orbital parameters of this ensemble serve to generate the time series of climatic precession and obliquity used in this work. Multitaper method spectral analysis is applied \citep[using \textit{Astrochron};][]{meyers2014astrochron} to the output of the obliquity model to extract the period of the main spin cycles. For each of the stable simulations, the spectral peak with the highest significant power is identified and their frequencies converted to the respective period. The results are 
summarized in Figures \ref{fig:precessionperiods} and \ref{fig:obliquityperiods}.

\begin{figure*}[ht!]
\plotone{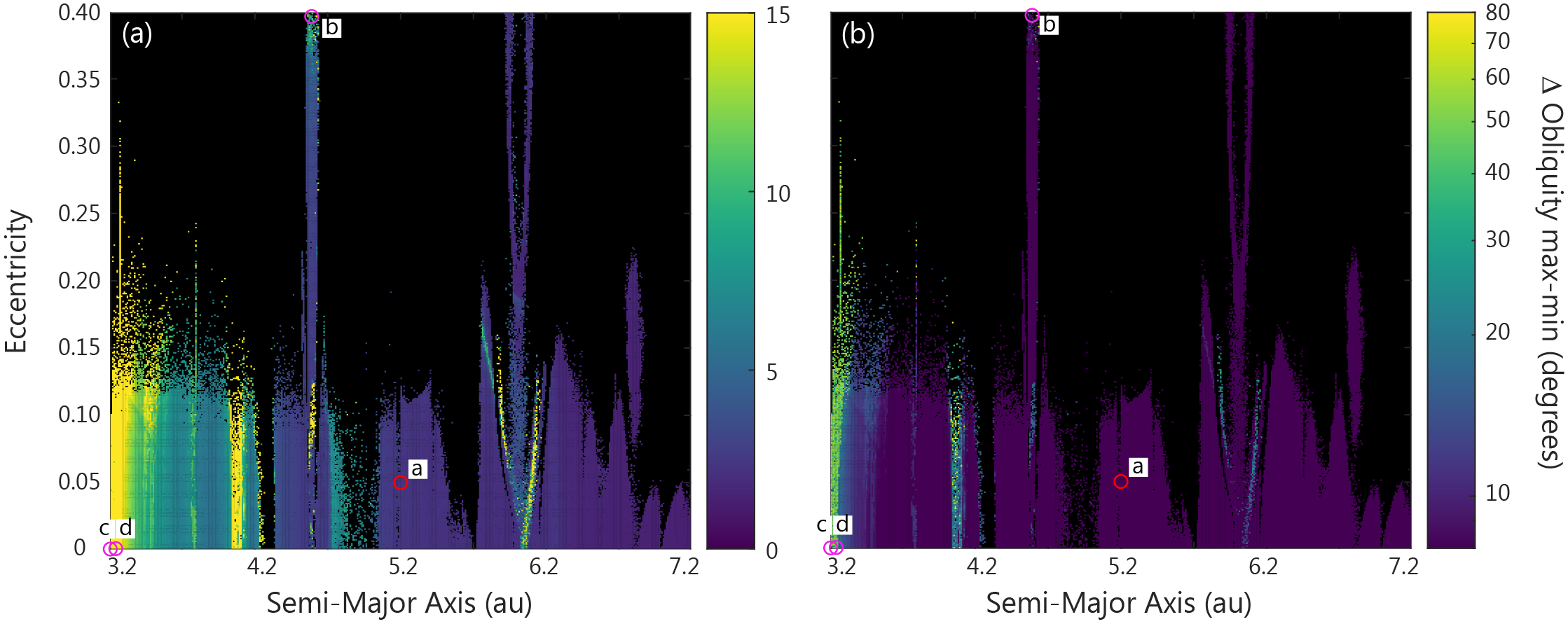}
\caption{The maximum variation in the obliquity (in degrees) of an Earth-like planet, under different planetary architectures, i.e. presence of a Jupiter-like planet with different initial semi-major axes and eccentricity. Both panels display the same data on (a) linear and (b) logarithmic scales to distinguish between high and low amplitude variations. The red circle tagged `a' indicates the current position of Jupiter in our Solar system. Pink circles tagged `b', `c', `d' indicate simulations used for the reconstruction of surface climate conditions in Section \ref{subsec:genieresults}.
\label{fig:obliquityamplitude}}
\end{figure*}

\subsection{Precession cycles} \label{subsec:precession}

\noindent The period of climatic precession cycles is sensitive to the architecture of a planetary system (Figure \ref{fig:precessionperiods}). In our ensemble of simulations, abrupt changes in the main period of precession cycles are attributed to a switch between the planet(s) with the strongest influence on the apsidal precession of the Earth-like planet.

Jupiter exerts a dominant control on the apsidal movement of Earth's orbit in our own Solar system, with a frequency g$_5$ of 4.3''~yr$^{-1}$, or period P$_{\omega 5}$ of 303~kyr, resulting in climatic precession cycles with a period of 23~kyr (Equation 2). If the Jupiter-mass planet is instead located at a smaller heliocentric distance, its apsidal precession cycles accelerate, and the frequency g$_5$ decreases. Whilst the apsidal precession of an Earth-like planet inherits some of this signal by precessing at a lower frequency, the impact on the climatic precession period is small, following Equation 2. When a Jupiter-mass planet is located between 3.2 and 5.3~au and has an eccentricity lower than approximately 0.05, the apsidal precession of Earth-like and Mars-like planets display resonant behaviour. The apsidal precession of the Earth-like planet varies rapidly with a frequency g$_3$ 86.4'' yr$^{-1}$ (or period of 15~kyr) while the apsidal precession of a Mars-like planet is about twice as rapid, varying with a frequency g$_4$ 178.7''~yr$^{-1}$ (or period of 7.3~kyr). The switch in the dominant resonant interaction is evident from the stark color transition that distinguishes dominant 24~kyr cycles driven by the Jupiter-mass planet from cycles with a period less than 17~kyr driven by the interaction between the Earth-like and Mars-like planets (Figure \ref{fig:precessionperiods}). 

The climatic precession period exceeds 30~kyr when the Jupiter-mass planet is located near 3.2~au with an eccentricity greater than 0.05, or when the Jupiter-mass planet is located near 4.6~au with an eccentricity close to 0.15 (yellow areas in Figure \ref{fig:precessionperiods}). Longer precession cycles near 3.2~au result from the interaction between the apsidal precession of the Earth-like, Mars-like and Jupiter-like planets with g$_5$ 1.82'' yr$^{-1}$. When $a_J\approx$ 4.6~au, the apsidal precession of the Earth-like, Venus-like, Jupiter-like, and Saturn-like planets are intimately connected, varying with frequency g$_5$ 7.52'' yr$^{-1}$.

\subsection{Obliquity cycles} \label{subsec:obliquity}

\noindent The obliquity period is related to the nodal precession frequency. With a Jupiter-mass planet closer to the star, the nodal precession cycles of the inner Earth-like planet shorten, resulting in obliquity cycles of lower frequency because the nodal and axial precession both rotate in a clockwise direction (Equation 1, Figure \ref{fig:obliquityperiods}, \ref{fig:climaticprecessionsketch}). The main axial obliquity cycles reach a period greater than 300~kyr when the Jupiter-mass planet is located near 3.2~au. Here, the nodal precession of the Earth-like and Mars-like planets display resonant behaviour with frequency s$_3$ of 43.03'' yr$^{-1}$ and frequency s$_4$ about twice as high. If the giant planet is positioned close to 6.1~au, the dominant obliquity cycles have a relatively low frequency. The orbital architecture of the planetary system here results in low amplitude 40~kyr obliquity cycles superimposed on chaotic, high amplitude cycles with a period greater than 1~Myr as a result of secular resonance between the Venus-like and Earth-like planet. 

The frequency of obliquity (and climatic precession) cycles have a critical influence on the surface climate evolution via time dependent climatic processes \citep[][]{spiegel2010,armstrong2004,deitrick2018a}. However, the amplitude of the cycles are equally important from a climatological perspective. The amplitude determines the variation in the amount of stellar energy received at any given latitude, both on seasonal and astronomical timescales. In the majority of our simulations, the obliquity does not vary by more than 5$\degr$ (Figure \ref{fig:obliquityamplitude}), consistent with the range of the simulated orbital inclinations \citep{horner2019}. More extreme obliquity oscillations occur at $e_J=$ 4.1~au, 4.5~au, and 4.7~au, where the high amplitude obliquity oscillations are the direct result of high amplitude oscillations in the orbital inclination. For instance, a 15$\degr$ range in the orbital inclination translates to a $2\times15\degr$~$\thickapprox$~30$\degr$ range in the obliquity (Figure \ref{fig:nodal}). 

\begin{figure*}[ht!]
\plotone{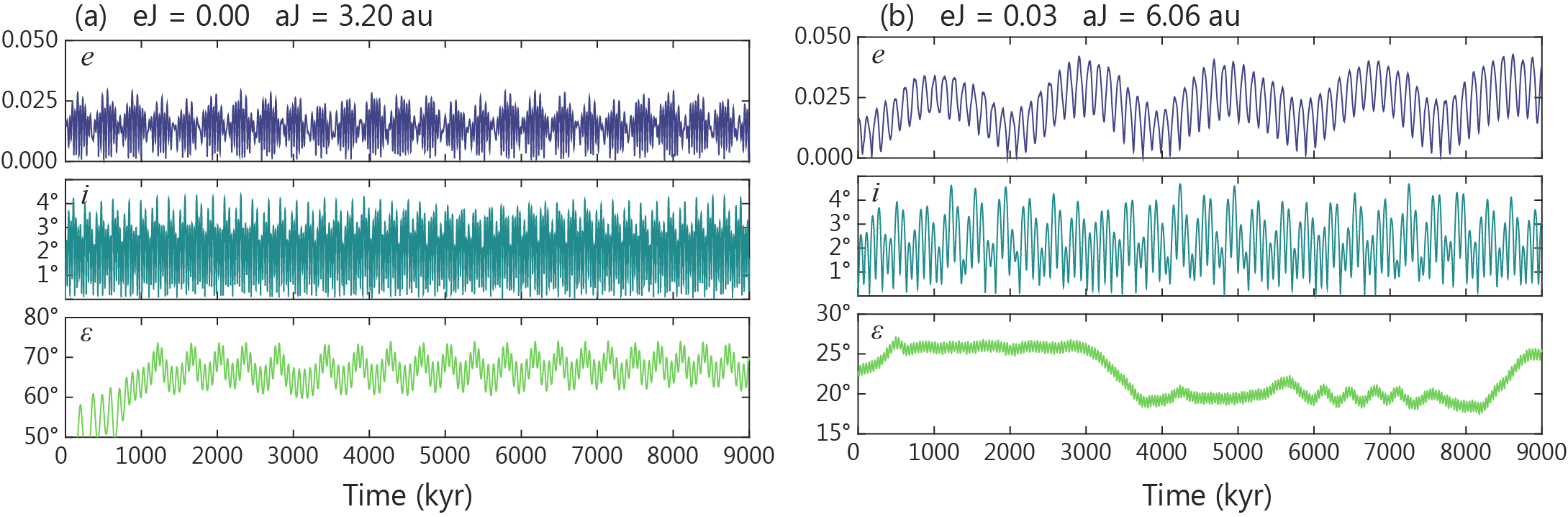}
\caption{Eccentricity $e$, inclination $i$, and obliquity $\epsilon$ cycles over nine million years for two example Earth-like planets. (a) High obliquity Earth-like planet that varies with an amplitude greater than the amplitude of orbital inclination. (b) Chaotic behaviour in the obliquity cycles of an Earth-like planet. Note the different y-axis scale for the bottom panels.
\label{fig:obliquityexamples}}
\end{figure*}

Some exceptionally high obliquity variations unrelated to orbital inclination are simulated when the Jupiter-mass planet is located at small semi-major axes (Figure \ref{fig:obliquityamplitude} and \ref{fig:obliquityexamples}a) and the period of a nodal precession cycle P$_\Omega$ is shorter than that of the axial precession cycle P$_\psi$ (Equation 2). The frequency of the nodal precession of the Earth-like planet is high enough to excite the axial obliquity to angles greater than those associated with the orbital inclination \citep{horner2019}. The resonant behaviour between the Venus-like and Earth-like planets excites the obliquity to values much greater than those of the orbital inclination in the region near 6.1~au (Figure \ref{fig:obliquityamplitude}). The obliquity evolution displays chaotic behaviour in these regions over the simulated timescales (Figure \ref{fig:obliquityexamples}b). 

\subsection{Surface climate simulations} \label{subsec:genieresults}

\begin{figure*}[ht!]
\epsscale{1.20}
\plotone{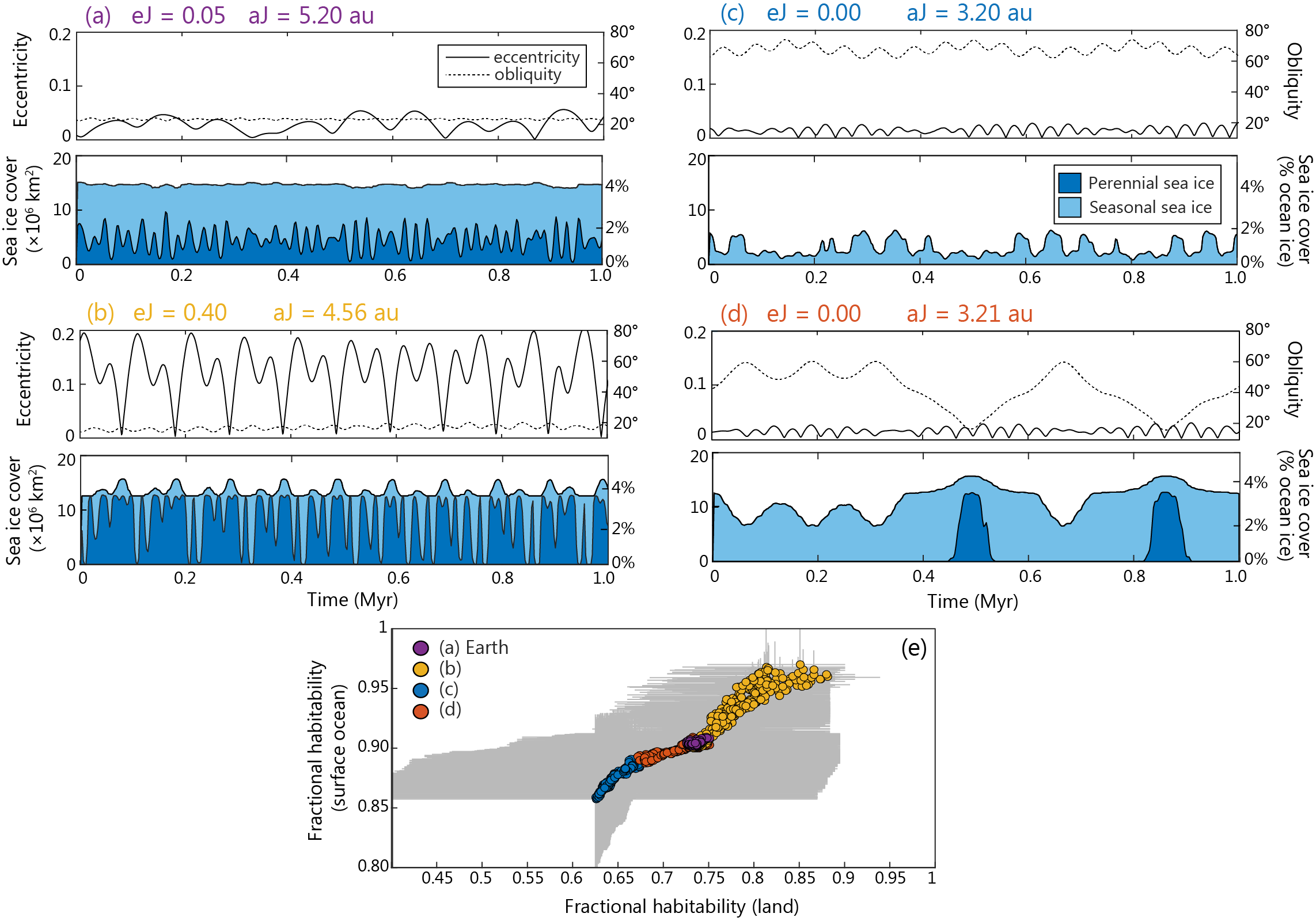}
\caption{Sea ice extent and fractional habitability on an Earth-like planet for four exemplar dynamically stable planetary systems. (a) Current Earth with Jupiter eccentricity $e_J$=0.05 and Jupiter semi-major axis $a_J$=5.20~au. (b) Highly eccentric Jupiter-mass planet $e_J$=0.40 and  $a_J$=4.56~au. (c) Close-in Jupiter-mass planet with circular orbit $e_J$=0.00 and  $a_J$=3.20~au. (d) Close-in Jupiter-mass planet with circular orbit  $e_J$=0.00 and  $a_J$=3.21~au. Eccentricity (solid) and obliquity (dashed) over time in the upper panels. The areal extent of perennial (year-round) and seasonal sea ice (in 10$^{6}$ km$^2$ and \% ocean area covered in ice) in the lower panels. (e) The annual mean fractional habitability over land versus ocean for each 2,500-year time interval of the four Earth-like planets. The grey area indicates the seasonal range in fractional habitability over land and ocean. 
\label{fig:genie}}
\end{figure*}

\noindent We select four examples to evaluate the impact of planetary architecture and orbital dynamics on the long-term climate conditions and habitability potential of an Earth-like planet. The four example cases, indicated in Figure \ref{fig:obliquityamplitude}, represent Earth-like planets that experience a wide range of orbital frequencies and amplitudes. Their calculated temporally-evolving eccentricity, precession, and obliquity parameters are applied to the Earth system model to calculate the total amount and spatial distribution of incoming stellar radiation. The model is integrated for 1~Myr (Figure \ref{fig:genie}). Two variables of interest for the purpose of discussing the impact of planetary architectures on the long-term climate evolution and habitability are the surface temperature and the areal extent of (sea) ice. Seasonally resolved 3D model output is generated every 2500 years by averaging the surface temperature and sea ice extent across three-month intervals to evaluate the changing surface conditions over astronomical timescales and the seasonal contrast under different astronomical configurations.

The method described by \cite{spiegel2009} is adopted to calculate the annual global mean fractional habitability for each saved time interval in all four simulations. A grid cell is deemed habitable and assigned a value of 1 if the overlying air temperature lies between 0-100$\degr$C. Grid cells covered in sea ice and/or with an overlying air temperature below 0$\degr$C or above 100$\degr$C, are assigned a value of 0 and are considered inhospitable. The annual global mean fractional habitability is spatially averaged and the seasonal range in fractional habitability is shown in grey (Figure \ref{fig:genie}e).

First, we simulate the evolution of Earth's surface conditions in our own Solar system (Figure \ref{fig:genie}a). The annual mean sea ice extent is about  11$\times10^{6}$~km$^2$ with minima during the Northern Hemisphere (NH) summer months of 4.5$\times10^{6}$~km$^2$ and maxima during NH winters of 15$\times10^{6}$~km$^2$  -- consistent with observational satellite data \citep{peng2018}. The extent of perennial (year-round) sea ice is largely controlled by the NH summer insolation driven by eccentricity-modulated precession forcing. Seasonal (winter) sea ice extends to an area about four times as large, and is less sensitive to precession and eccentricity. The amount of irradiation that reaches polar latitudes during winter is minimal and partially controlled by the obliquity angle. Across the 1 Myr simulation, the annual mean fractional habitability varies only minimally. Approximately 75\% of land area is considered habitable whereas the fractional habitability of the surface ocean is about 90\% (Figure \ref{fig:genie}e), resulting in a weighted global mean fractional habitability of 85\%. This value is identical to that of the fractional habitability inferred from the 2004 NCEP/NCAR temperature reconstructions \citep{spiegel2009}.

The second example demonstrates the climatic effect of relatively short period and high amplitude eccentricity and precession cycles using a simulation with a Jupiter-mass planet located at 4.65~au and an eccentricity of 0.40. The two main eccentricity cycles experienced by the Earth-like planet have periods of 50 and 100~kyr varying over a total range of 0.2. Precession cycles have a 25~kyr period (Figure \ref{fig:precessionperiods}). The extent of seasonal sea ice is directly controlled by eccentricity forcing. A more eccentric orbit results in elevated global mean temperatures and reduced polar sea ice, particularly during the NH summer months. The extent of perennial sea ice varies more rapidly with precession forcing modulated by the eccentricity (Figure \ref{fig:genie}b). The low obliquity angle minimizes seasonal contrast and allows for a relatively greater maximum extent of perennial sea ice (13$\times$10$^{6}$~km$^2$) when the NH is tilted toward the star during aphelion, regardless of the orbital eccentricity of the Earth-like planet at the time. Despite the greater annual mean extent of sea ice in this simulation compared to that of the modern Earth, the fractional habitability of the surface ocean is greater (Figure \ref{fig:genie}e). For the Earth-like planets in this study, the fractional habitability is determined mainly by the lower 0$\degr$C limit rather than the upper 100$\degr$C limit. Even at high eccentricity, the thermal inertia of the ocean maintains temperatures well below 100$\degr$C. However, large parts of the sea ice-free Southern Ocean that experience subzero temperatures on the modern Earth reach temperatures above 0$\degr$C when the eccentricity is higher and obliquity lower. This raises the fractional habitability. The fractional habitability over land is also higher compared to the modern Earth simulation because the annual global mean stellar energy received at the top of the atmosphere increases with eccentricity, resulting in higher overall temperatures and a reduced extent of regions with temperatures below 0$\degr$C. This scenario has the smallest seasonal range in fractional habitability due to the low obliquity (Figure \ref{fig:genie}e).

The smallest areal extent of sea ice is simulated in the third scenario considered, where a Jupiter-mass planet is positioned at 3.20~au. The mean obliquity angle across the 1~Myr simulation is approximately 70$\degr$ (Figure \ref{fig:genie}c). All sea ice melts during summer but winter temperatures drop below freezing, facilitating the production of seasonal sea ice with a maximum area of 7$\times10^{6}$~km$^2$ when the obliquity angle is smallest. The fractional habitability over land and ocean is much reduced compared to Earth today (Figure \ref{fig:genie}e) because the extreme obliquity drives winter temperatures below 0$\degr$C across a wide range of latitudes. The minimum fractional habitability over land and ocean is 37\% and 80\%, respectively. 

High amplitude variations in the obliquity angle occur when a Jupiter-mass planet is located at 3.21~au (Figure \ref{fig:genie}d). The $\sim$60$\degr$ obliquity angle prevents the formation and preservation of sea ice throughout the summer months so sea ice is only present during winter in the first 400~kyr. When the obliquity falls below $\sim$30$\degr$, the polar summer insolation becomes sufficiently low to preserve perennial sea ice. In this scenario, the eccentricity variations in the orbit of the Earth-like planet are too small to exert a notable effect on the surface conditions. The mean fractional habitability is slightly lower than that of our Earth due to the periodically high obliquity and extreme seasonality that cause an increase in the area experiencing sub-freezing temperatures during winter (Figure \ref{fig:genie}e).

A recurring feature in all simulations is that the fractional habitability over the surface ocean is typically greater than over land. The large heat capacity of the ocean helps to maintain relatively warm temperatures throughout the winter and mutes seasonal variability.

\begin{figure}[]
\epsscale{1.0}
\plotone{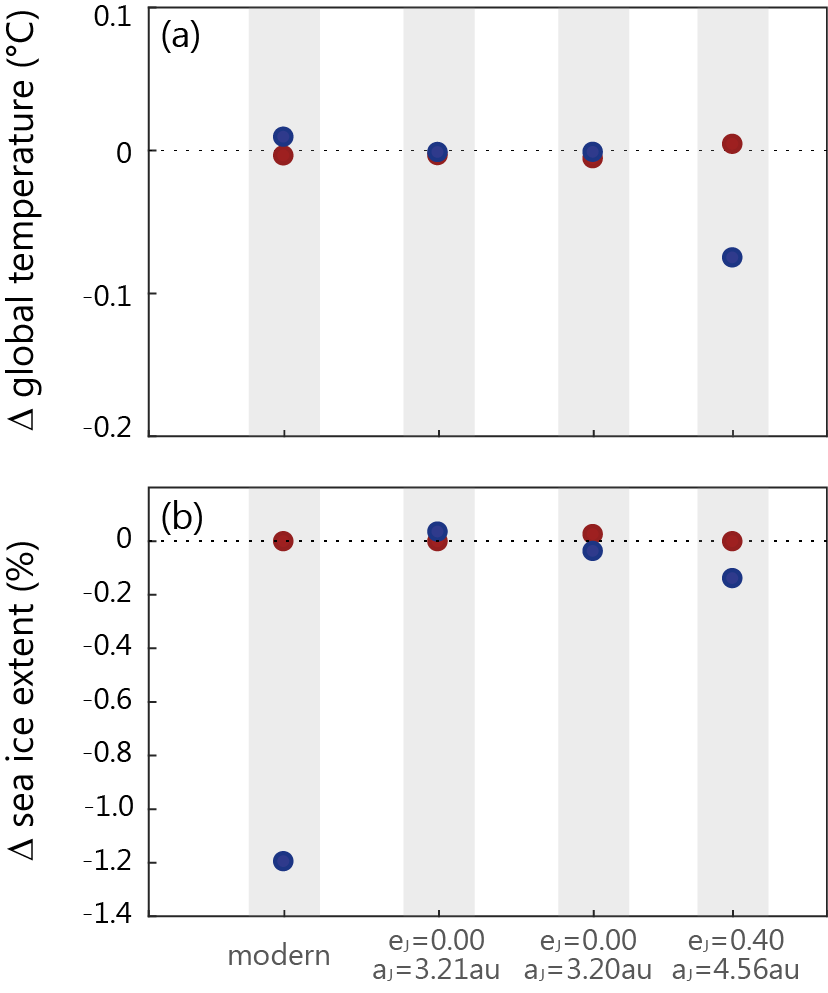}
\caption{The difference in (a) the annual global mean air temperature (in $\degr$C) and (b) the annual mean sea ice extent (in \%) between eight steady state runs and transient snapshots corresponding to the four examples in Figure \ref{fig:genie}.
\label{fig:genie_anom}}
\end{figure}

In addition to the transient simulations, two steady state runs were carried out for each of the four examples, with fixed astronomical parameters corresponding to certain time intervals. Steady state simulations were run for a total of 20~kyr to ensure equilibrium surface conditions. By comparing the steady state conditions to those of the transient run at corresponding time intervals, we quantify the degree to which time dependent climate processes respond to varying astronomical forcing. In all examples, the differences between the transient snapshots and steady state simulations are small (Figure \ref{fig:genie_anom}). The annual global mean air temperature does not differ by more than 0.08$\degr$C while the areal sea ice extent in steady state simulations does not deviate more than 1.2$\%$ from that simulated in transient runs.

However, differences between the transient versus steady state conditions should be considered a minimum. The only processes providing climatic inertia in our climate model set-up are sea ice growth and deep ocean circulation that also control the model equilibrium timescale. Inclusion of additional slowly responding components of the Earth system, such as land-based ice or much larger ocean volume, could drive greater differences between transient and steady state experiments. Our results are consistent with the comparison made in \cite{lunt2006}, who found good agreement between equilibrium and transient simulations in the ocean-atmosphere system. 

\section{Implications for long-term habitability potential} 
\label{sec:climate}

\noindent Geological archives on Earth contain ample evidence that small variations in astronomical forcing contribute to global climate variations, most famously the extensive glacial-interglacial cycles of the Quaternary period linked to the gravitational dynamics between Jupiter and the inner terrestrial planets \citep{milankovitch1941,hays1976,imbrie1992}. This leads us to wonder about the impact of astronomical cycles on surface climate variability and habitability in planetary systems with different architectures. Our \textit{n}-body, obliquity, and million-year-long climate simulations suggest that extreme eccentricity cycles, such as those in systems with a highly eccentric giant companion planet, increase the fractional habitability (Figure \ref{fig:genie}e) while also preventing global freezing of planets at the outer edge of the HZ \citep{spiegel2010,dressing2010}. In contrast, extreme obliquity variations such as those in systems with a giant planet closer inward (Figure \ref{fig:genie}cd) can also expand the outer edge of the HZ \citep{armstrong2014} but simultaneously reduce the fractional habitability (Figure \ref{fig:genie}e) as a greater surface area experiences extremely cold or hot temperatures when the planet is (temporarily) highly tilted relative to the orbital plane. Hence, to assess the long-term habitability of a planet, eccentricity variations should ideally be considered in tandem with potential obliquity (and precession) cycles. 
 
In none of the examples does the fractional habitability drop to zero, implying that the surface remains at least partially habitable during the four 1~Myr simulated intervals. It should, however, be noted that the variability simulated here is likely an underestimation for the following reasons. First, our model lacks continental ice sheets and snow coverage, and only simulates thermodynamically formed sea ice. To account for the missing albedo effect of land-based ice and snow, we imposed a fixed albedo over the polar latitudes, thereby omitting any snow and ice-albedo feedbacks over land that might alter the climate response to astronomical forcing. It is not necessarily obvious that land-based ice and sea ice respond similarly to astronomical forcing. Continental snow and ice caps have a higher albedo \citep{perovich1986,key2001}, thereby making them more resilient against melting and able to grow more rapidly. In addition, the wavelengths at which these surfaces reflect \citep{joshi2012,shields2013} and their dynamical controls \citep{pollard1978,weaver2001} are different \edit1{from} sea ice. Continental ice/snow volume is driven mainly by atmospheric conditions and internal feedbacks but sea ice volume is additionally driven by the thermodynamics of the underlying ocean. What this implies for the rate at which ice sheets respond to isolation forcing remains uncertain \citep{bamber2007}. 

Secondly, a greater climate sensitivity to astronomical forcing can result from additional (positive) feedbacks related to physical, chemical, and biological processes. For example, during glacial-interglacial intervals, the cycling and storage of carbon in the ocean and atmosphere changed in response to astronomical forcing. During cold glacials, more carbon is stored in the ocean and less carbon resides in the atmosphere, which subsequently lowers the CO$_2$ radiative forcing and thereby cools global climate further \citep{sigman2000,kohfeld2009}. Such positive feedback loops amplify the astronomical climate variability though negative feedbacks also exist. The net impact is uncertain on exoplanets that likely experience physical, chemical, and potentially biological processes of different magnitudes and timescales.

Finally, the simplified physics in the atmosphere model likely reduces the simulated astronomical climate variability, particularly when considering the impact of astronomical cycles that deviate notably from the modern Earth in both amplitude and frequency. For instance, the obliquity angle has great control over the meridional heat transport \citep{spiegel2009,ferreira2014,linsenmeier2015,kilic2017}. Whilst our model accounts for diffusive atmospheric and oceanic heat transport, the wind patterns (affecting sensible heat and moisture transport) are fixed and unresponsive to astronomical forcing. When obliquity reaches a critical threshold of 54$\degr$, the polar regions receive more insolation than the equatorial regions on an annual mean basis and thus heat is net transported from the high to low latitudes which would cause a dramatic reorganization of atmospheric dynamics \citep{williams2003,ferreira2014,linsenmeier2015}. Likewise, we do not account for changes in the cloud coverage and cloud feedbacks. Both could behave very differently on planets spinning at highly oblique angles \citep{kang2019}.

Whilst considering the above caveats, the surface climate remains habitable in all four examples, with temperate conditions and a large ice-free area across the 1~Myr simulation time. In the context of habitability and the ability of any life forms to survive at the surface, we must also consider the interannual variability. The most extreme three-month average seasonal contrast ($\sim$80$\degr$C) occurs above land under the highest obliquity angles, compared to a maximum seasonal contrast of $\sim$38$\degr$C in the simulation most comparable to the modern Earth. Studies of the impact of climate change on ectotherms have shown that species living at the highest latitudes on Earth are generally more resilient to temperature fluctuations as they evolved to have a broader thermal tolerance to survive seasonal temperature fluctuations \citep{addo2000,Deutsch2008}. A high seasonal temperature contrast is thus not necessarily deleterious for life as we know it. However, organisms that evolved under moderate seasonal temperature fluctuation (such as those living in tropical regions on Earth) have a much narrower thermal tolerance range and are much more susceptible to change \citep{Deutsch2008}. This may pose a problem in the fourth example scenario (Figure \ref{fig:genie}d) when the obliquity angle fluctuates dramatically and thereby surface conditions experience cycles of sedate-to-extreme seasonal contrast. In terms of habitability, it is important to assess the seasonal surface variability, but also the variability on astronomical timescales.

The response time of the surface conditions (climate inertia) to climate perturbations comes into play when the radiative balance is rapidly perturbed, as would happen when the distribution of incidence stellar flux is continuously altered through high frequency astronomical cycles \citep[e.g.][]{armstrong2014,georgakarakos2018}. The surface system may not be able to re-equilibrate fully under the influence of rapid oscillations and may subsequently experience a weakened response to astronomical forcing. An example is the suppressed positive ice-albedo feedback during rapid obliquity cycles \citep{armstrong2014}. The transient non-steady state simulations presented here take into account the more rapid ocean-sea ice feedbacks, but the period of even the shortest simulated astronomical cycles is long enough for sea ice to equilibrate to the new radiative conditions (Figure \ref{fig:genie_anom}). More rapid oscillations than those simulated here are required to suppress the (sea) ice-albedo feedback, although we note that the presence of dynamic continental ice sheets (not modeled here) may change the ice-albedo response time.

\begin{figure*}[ht!]
\plotone{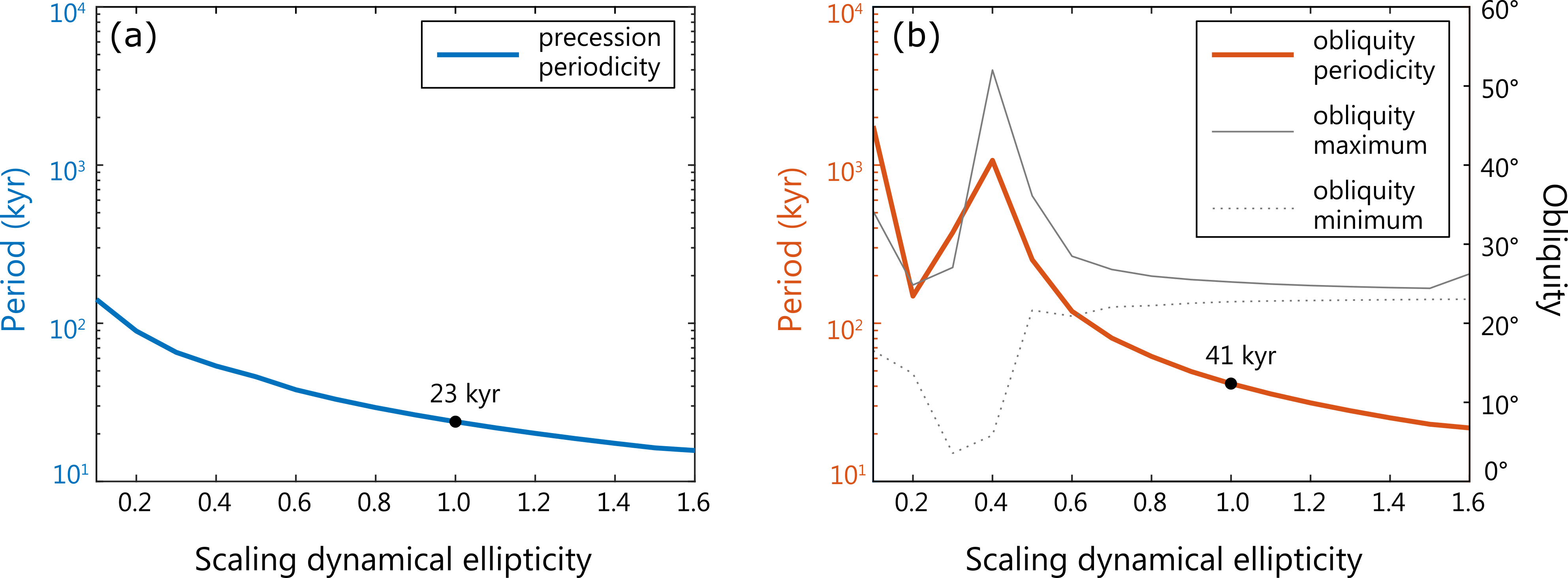}
\caption{Duration of climatic precession (left panel) and obliquity (right panel) cycles as a function of the dynamical ellipticity scaling factor, following \cite{laskar1993}, for simulation where Jupiter is located at it's current position with an eccentricity of 0.05. A scaling factor of 1.0 equates to a dynamical ellipticity of 0.00328 equal to the value for the modern Earth.
\label{fig:ellipticity}}
\end{figure*}

\section{Application to HZ planets} \label{sec:discussion}

\noindent For the majority of the dynamical simulations presented here, the amplitude and frequency of obliquity cycles can be estimated directly from the orbital elements $i$, $\Omega$, and $\omega$, in combination with the parameter that describes the axial precession, $\psi$. The time evolution of the orbital elements of exoplanets in known systems can thus be modelled with $n$-body simulations and provide a first order estimate for the frequency and amplitude at which the obliquity varies \citep{atobe2004,armstrong2014,shan2018}, as has recently been done for Kepler-62f \citep{shields2016,quarles2020}. Such simulations also reveal whether the spin motions of a terrestrial planet may be subjected to intense perturbations due to resonant behavior, and are therefore a useful tool for exoplanet research toward reconstructing possible climate states of terrestrial exoplanets in the Habitable Zone.

The major unknowns required to reconstruct the spin dynamics of planets beyond the Solar system are related to the precession rate of the rotational axis (Equation 3 and A.6). This motion depends on the dynamical ellipticity (proportional to the planetary rotation rate for rapidly rotating planets), tidal torques, and the initial obliquity of the planet. These parameters are currently unobtainable for exoplanets or can only be roughly estimated at best. Below, we evaluate how tidal torques and dynamical ellipticity influence the simulation of precession and obliquity cycles using the following equation: 

\begin{equation}
\frac{d\psi}{dt} = \frac{3}{2} \left( \frac{M_{m}n_m^2}{\nu} +  \frac{M_{s}n_s^2}{\nu} \right) E_D \cos \epsilon
\end{equation}

\noindent where  $M_{m}$ is the ratio between the mass of the Moon and the summed mass of the Moon and Earth, and $n_m$ is the mean motion of the Moon (2$\pi$/27.3 days). $M_{s}$ is the ratio between the mass of the Sun and the summed mass of the Sun and Earth ($\thickapprox$ 1). The mean motion of the Earth is $n_m$ (2$\pi$/365.25 days). Finally, $\nu$ is Earth's angular momentum, $E_D$ is the dynamical ellipticity, and $\epsilon$ is the axial obliquity. 

\subsection{Tidal torques} \label{subsec:moon}

\noindent The Earth may be a rare case in that our planet is accompanied by a relatively large moon. The benefits of having a moon in terms of planetary habitability and the origin of life have been discussed in various studies \citep[e.g.][]{laskar1993moon,lathe2004,lissauer2012}. Because many terrestrial exoplanets may not have a massive satellite \citep{elser2011}, we evaluate how the frequency of axial precession and obliquity cycles are affected in the absence of lunar torques. Following Equation 3, Earth's axial precession rotates at 6.744$\times10^{-7}$~rad per day, when accounting for tidal contributions from both the Moon and Sun. The axial precession rate decreases to 2.124$\times10^{-7}$~rad per day when tidal torques from the Moon are eliminated from the equation, resulting in an axial precession cycle with a approximately 81~kyr period. The resulting obliquity cycles would have a period of 424~kyr (Equation 1), and climatic precession would oscillate with a period of 64~kyr (Equation 2). Even though various studies have shown that the amplitude of the obliquity cycles would increase in absence of a Moon \citep{laskar1993moon,lissauer2012,li2014}, the period of those cycles would become significantly longer, making the rate of change experienced by the surface relatively more sedate. 

Other objects in the Solar system, including Jupiter, are either too distant or have a mass too small to induce significant tidal effects on Earth. However, tidal torques scale with distance and it is worth investigating whether a giant planet would exert a notable torque $\tau_{Jup}$ on the equatorial bulge of an Earth-like planet if it was positioned in closer proximity. We use the following relationship:

\begin{equation}
\tau \propto \frac{m}{a^{3}}
\end{equation}

\noindent where $m$ is the mass of the object exerting the torque in solar masses ($M_{\Sun}$) and \textit{a} is the average distance between the object and Earth in astronomical units ($au$). The torque of the Moon $\tau_{moon} \propto$ 2.2 is about twice as strong as the tidal torque from the Sun $\tau_{sun} \propto$ 1. With Jupiter located at its current semi-major axis of 5.2 au, $\tau_{Jup} \propto$ 0.00018, the torque is negligible compared to the solar and lunar torques. Jupiter at 3.2 au results in $\tau_{Jup} \propto$ 0.00030 and thus remains a negligible force on the planet's equatorial bulge in comparison.

\subsection{Dynamical ellipticity} \label{subsec:ellipticity}

\noindent The dynamical ellipticity E$_D$ is a measure of the oblateness of a planet partly caused by centrifugal forces associated with the planetary rotation, and to a lesser extent by the geography, topography \citep{ward1979}, presence and location of massive ice caps \citep{dehant1990,rubincam1990,ito1995}, core-mantle dynamics \citep{neron1997,forte1997,correia2006}, and atmospheric thickness \citep{barnes1983,volland1996}. The torques on a more oblate planet with a larger equatorial bulge are stronger compared to the torques on a more spherical planet, resulting in a more rapid precession of the rotational axis (Figure \ref{fig:ellipticity}). Using a parameterization for E$_D$ in the obliquity model used here \citep{laskar1993}, we simulate how this impacts the frequency of climatic precession and obliquity cycles.

Increasing the E$_D$ scaling factor to 1.6 shortens the obliquity period to about 20 kyr whilst a more rigid Earth with a dynamical ellipticity scaling factor of 0.6 lengthens obliquity cycles to 119 kyr. For comparison, the dynamical ellipticity of Mars is estimated at E$_D$ = 0.0054 (1.6$\times$Earth) \citep{bouquillon1999} and that of Venus is likely much smaller, E$_D$ = 0.000013 due to its slow rotation rate \citep{yoder1995,correia2003a}. Considering how sensitive the frequency of astronomical cycles are to the dynamical ellipticity, it will be essential to determine the rotation rates of HZ exoplanets before being able to discuss the long-term habitability and stability of the surface conditions on geological and astronomical timescales.

\section{Conclusion} \label{sec:conclusion}

\noindent Surface conditions and fractional habitability of (exo)planets in the HZ vary over time as a function of the architecture of their planetary system and orbital dynamics. The frequency and amplitude of eccentricity and orbital inclination cycles of an Earth-like planet are sensitive to the relatively minor variations in the orbital characteristics of companion planets \citep{horner2019}. These in turn drive the magnitude and frequency of axial obliquity and precession cycles with implications for surface climate conditions and long-term habitability. This study is the first to apply \textit{n}-body and obliquity model output to a 3D ocean-coupled climate model where the eccentricity, obliquity, and precession vary transiently on the million-year timescales relevant for the evolutionary development of life.

Even the more extreme orbital cycles presented here result in (partially) habitable surface conditions. The large heat capacity of the vast ocean basin maintains temperate ocean temperatures although the fractional habitability over land varies strongly seasonally and annually on astronomical timescales. A planet is less likely to be suitable for life when the fractional habitability drops, permanently or temporarily, to near zero values, which may happen for exoplanets closer to the edge of the HZ and/or with low thermal inertia in the climate system while experiencing extreme astronomical cycles.

Our results support previous work that demonstrates how eccentricity and obliquity cycling can act to inhibit large-scale freezing with a sufficiently large heat reservoir, and thereby expand the outer edge of the HZ. At the same time, high mean obliquity or extreme obliquity cycles that arise when a giant planet is close-in, can act to reduce the habitable surface area. In contrast, the habitable surface area increases under extreme eccentricity cycles that arise under the influence of an eccentric giant planet. An eccentric giant companion may therefore be favorable for the habitability of a smaller terrestrial planet, provided its distance is large enough from the smaller planet to prevent high-amplitude obliquity cycles.

Transient ocean-climate simulations differ from conventional steady state models because slower climate processes, such as the growth of sea ice and heat storage in the vast ocean basin, prevent the system from reaching full equilibrium with respect to changing orbital parameters. However, even the most rapid astronomical changes investigated here (50 kyr eccentricity cycles) are sufficiently slow for sea ice and ocean heat storage to approach an equilibrium state. A planet with smaller heat capacity or more rapid changes to the radiative balance may not be buffered against rapid climate variability.

A similar sequential methodology of using $n$-body simulations, obliquity models, combined with transient ocean-coupled climate simulations can be applied to assess the long-term habitability potential of HZ exoplanets in planetary systems provided that the planetary masses and orbital architecture are relatively well constrained. A series of $n$-body simulations produces possible dynamical evolution pathways of the orbital elements (\textit{i}, $\Omega$, $\omega$) that give insight into the dynamical stability of the system and allow a first-order estimate of the potential amplitude and frequency of obliquity cycles using educated assumptions about the rate of axial precession.

The era of exoplanet exploration and characterization has only just begun. Many of the parameters required to reliably simulate orbital and spin dynamics of HZ exoplanets such as the planetary rotation rate, initial obliquity, axial precession rate, or dynamical ellipticity are currently unobtainable. However, it is expected that significant advances will be made in observational facilities and analysis techniques in the coming years. This, combined with our growing understanding of the evolution of protoplanetary disks and planet formation, will undoubtedly provide us with a plethora of new tools to improve exoplanet characterization, and provide the critical data needed to more reliably assess the potential habitability of newly discovered exoplanets to help focus the future search for evidence of life beyond the Earth to the most promising targets.

\acknowledgments
\noindent While working on this manuscript, PV and SKT were supported by a Heising-Simons Foundation award. The results reported herein benefited from collaborations and/or information exchange within NASA's Nexus for Exoplanet System Science (NExSS) research coordination network sponsored by NASA's Science Mission Directorate.


\software {Mercury} \citep{chambers1999}; Astrochron R package \citep{meyers2014astrochron}; La1993 code \citep{laskar1993} accessible via \url{http://vizier.cfa.harvard.edu/ftp/cats/VI/63/}
          
\newpage


\appendix
\renewcommand{\theequation}{A.\arabic{equation}}
\renewcommand\thefigure{A.\arabic{figure}}
\setcounter{figure}{0} 

\noindent The obliquity model described in \cite{laskar1993} is applied for the integration of Earth's obliquity ($\epsilon$) and precession (\textit{p$_A$}) over time for each of the stable dynamical $n$-body simulations in \cite{horner2019}. The obliquity model follows the following equations based on the rigid Earth theory of \cite{kinoshita1977}.

\begin{equation}
\frac{dp_A}{dt} = R(\epsilon) - \textrm{cot } \epsilon [  A(p,q)\textrm{sin }p_A + B(p,q) \textrm{cos } p_A] - 2C(p,q)- p_g
\end{equation}
\begin{equation}
\frac{d\epsilon}{dt} = -B(p,q)\textrm{sin }p_A + A(p,q) \textrm{cos } p_A
\end{equation}

with:

\begin{equation}
A(p,q) = \frac{2}{\sqrt{1-p^2-q^2}}(\dot{q}+p(q\dot{p}-p\dot{q}))
\end{equation}
\begin{equation}
B(p,q) = \frac{2}{\sqrt{1-p^2-q^2}}(\dot{p}-q(q\dot{p}-p\dot{q}))
\end{equation}
\begin{equation}
C(p,q) = (q\dot{p}-p\dot{q})
\end{equation}

and:

\begin{equation}
\begin{aligned}
R(\epsilon) = & \frac{3k^2m_M}{a^3_M\nu}\frac{C-A}{C} \\
& \times \left[(M_0-M_2/2)\textrm{cos }\epsilon + M_1\frac{\textrm{cos }2\epsilon}{\textrm{sin }\epsilon} - M_3\frac{m_M}{m_E+m_M}\frac{n^2_M}{\nu n_\Omega}\frac{C-A}{C}(6\textrm{cos}^2\epsilon -1)\right] \\
& + \frac{3k^2m_\sun}{a^3_\sun\nu}\frac{C-A}{C}[S_0 \textrm{cos }\epsilon]
\end{aligned}
\end{equation}

\noindent where p = sin(\textit{i}/2)sin($\Omega$) and q = sin(\textit{i}/2)cos($\Omega$). Numerical values for each parameter are given in \citep[]{laskar1993}. All parameters are set as their original values that were used to reconstruct the rotational evolution of our modern Earth for direct comparability between cycles experienced by modern Earth and the alternative Earth-like planets simulated in this study.

\begin{figure}[h]
\plotone{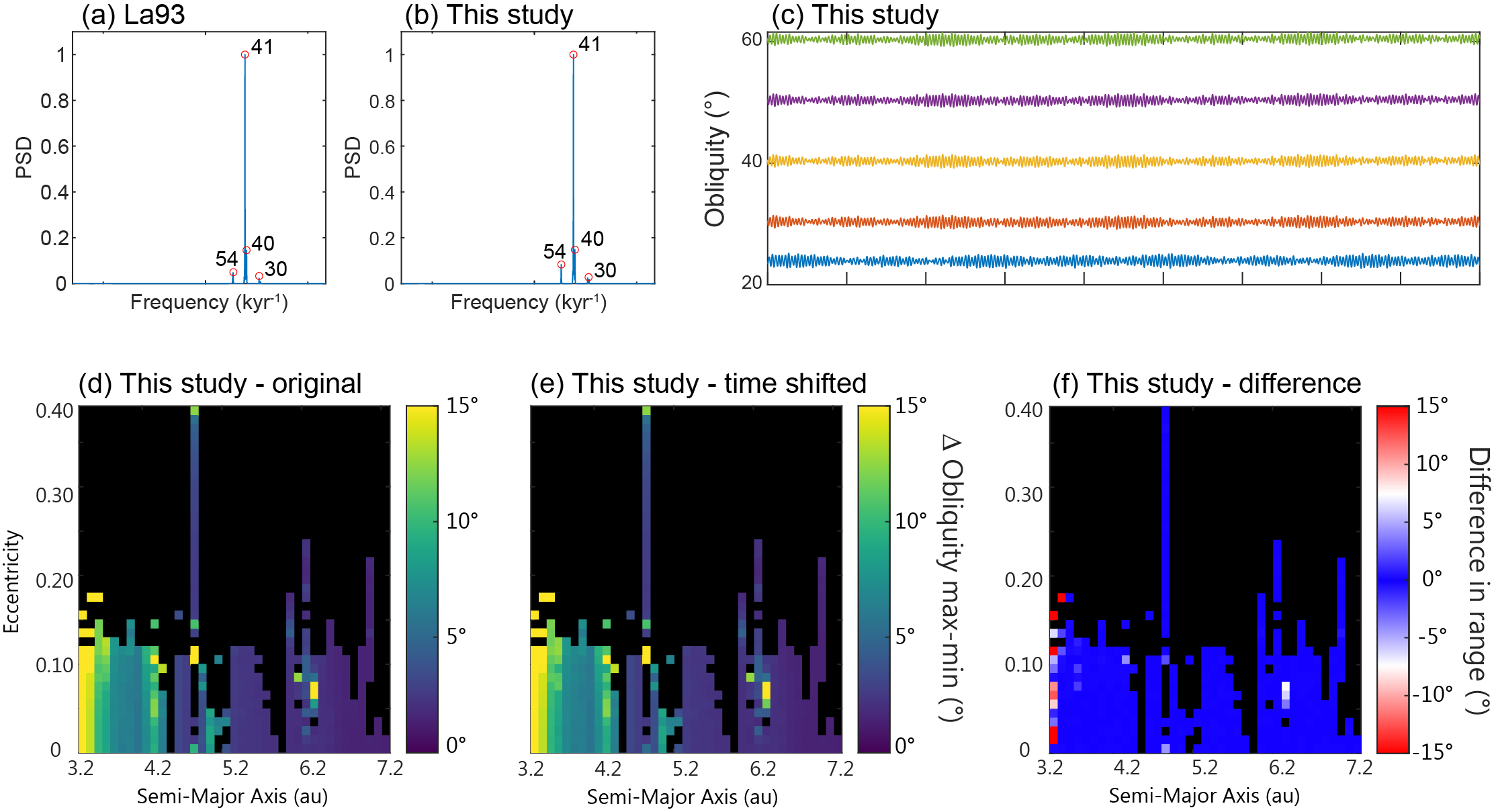}
\caption{Testing the obliquity model. (a) Fast Fourier power spectrum of the La93 astronomical solution. (b) Fast Fourier power spectrum of the simulation in this study most comparable to the Solar system configuration. (c) Obliquity model output for initial obliquity values of 23\degr, 30\degr, 40\degr, 50\degr, 60\degr for the modern Solar system configuration. (d) The maximum variation in the obliquity (in degrees) of an Earth-like planet under different planetary architectures, i.e. Figure \ref{fig:obliquityexamples}a in low resolution (40$\times$40~simulations across the full parameter space). (e) As Figure \ref{fig:test_model}d but input signal shifted in time by 100 year to demonstrate sensitivity to initial conditions. (f) Difference in maximum obliquity variation between \ref{fig:test_model}e and \ref{fig:test_model}d.
\label{fig:test_model}}
\end{figure}

Several tests are performed to verify the results. First, the simulation most comparable to the modern Solar System, with Jupiter at 5.2~au and an eccentricity of 0.05, is compared to the existing (La93) astronomical solution. Both amplitudes and power spectra of precession and obliquity cycles match with the La93 \citep{laskar1993} solution (Figure \ref{fig:test_model}ab). Secondly, we test the sensitivity to the initial obliquity angle for the modern. Initial values of $\epsilon_0$=23\degr, $\epsilon_0$=30\degr, $\epsilon_0$=40\degr, $\epsilon_0$=50\degr, $\epsilon_0$=60\degr are used for integration. Despite the very different simulated mean obliquity, the amplitudes and main periods remain unchanged (Figure \ref{fig:test_model}c).

Some simulations display an initial drift in obliquity across the first few years (e.g. Figure \ref{fig:obliquityexamples}a). To verify that this drift does not result from sensitivity of the model equations to small deviations, we run sensitivity tests for a small sub-sample ($40\times40$) of simulations in which the input signal is slightly offset. For instance, the original input signal has a time step of 1000 year, i.e. t~=~0, 1000, 2000, etc. In the sensitivity runs, the input signal is shifted by 100 years, i.e. t~=~100, 1100, 2100, etc. The differences at the start of the simulations that show an initial drift are negligible and confirm the transient nature of the first few thousands of years while the planet adjusts to the new conditions. To prevent the transient drift at the start of the simulation from distorting the results of the amplitude calculations, the first million year is removed from the analysis when extracting the maximum obliquity amplitude (Figure \ref{fig:obliquityamplitude}). The sensitivity test also reveals that the model is sensitive to the initial conditions (Figure \ref{fig:test_model}d-f). Whilst the majority of simulations in the low-resolution parameter space appear to have a low sensitivity to initial conditions in terms of the maximum obliquity variation, those regions where the Jupiter-like companion is close to 3.2~au display a greater deviation between the two sensitivity simulations and should be interpreted with caution.

\begin{figure}[ht!]
\plotone{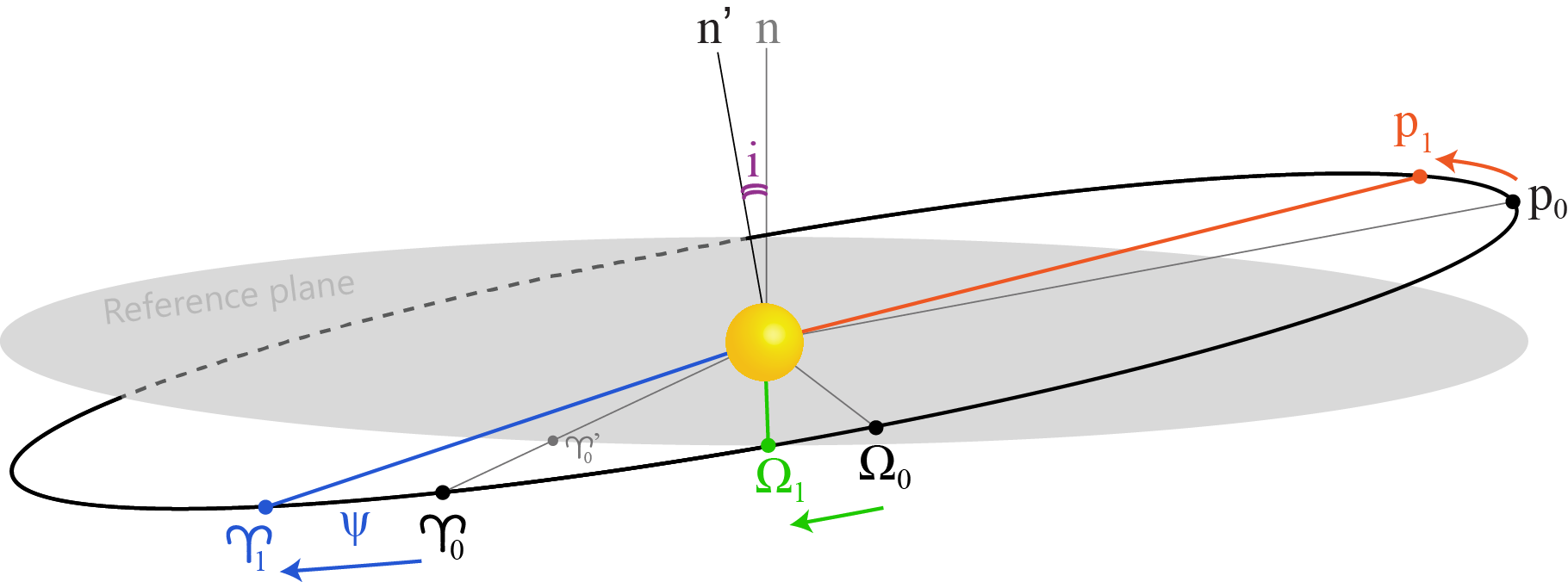}
\caption{Schematic depiction of the relative positions of the ascending node ($\Omega_0$ and $\Omega_1$), perihelion ($p_0$ and $p_1$), and the moving vernal equinox ($\Upsilon_0$ and $\Upsilon_1$) at two sequential points in time, controlled by the clockwise movement of axial precession ($\Psi$) on the Earth-like planet. Their relative direction of movement is indicated with arrows. $N$-body simulations provide the angle (longitude) of the ascending node relative to a reference point ($\Upsilon'_0$) on the reference plane with orbit normal ($n$). The reference point is the first point of Aries during the year 2000. The orbital inclination ($i$) is the angle between the plane of reference and orbital plane with normal ($n'$). 
\label{fig:climaticprecessionsketch}}
\end{figure}




\end{document}